\documentclass{article}
\usepackage{spconf,amsmath,graphicx}

\usepackage[english]{babel}

\usepackage{amsmath}
\usepackage{amsfonts}
\usepackage{graphicx}
\usepackage[acronym]{glossaries}
\usepackage[colorlinks=true, allcolors=blue]{hyperref}
\usepackage[font=footnotesize]{subfig}
\usepackage{cite}
\usepackage{float}
\usepackage{multirow}

\usepackage{tikz}
\usetikzlibrary{matrix,chains,positioning,decorations.pathreplacing,arrows,arrows.meta,external,shapes, shapes.geometric,shapes.misc,calc,external,patterns}

\newacronym{asr}{ASR}{Automatic Speech Recognition}
\newacronym{wer}{WER}{Word Error Rate}
\newacronym{etoe}{E2E}{End-to-end}
\newacronym{lm}{LM}{Language Model}
\newacronym{lstm}{LSTM}{Long Short-Term Memory}
\newacronym{bilstm}{BiLSTM}{Bidirectional Long Short-Term Memory}

\newcommand{\ie}{\textit{i.e.~}}

\newcommand\ignore[1]{}

\newcommand{\figref}[1]{Fig.~\ref{#1}}

\newcommand{\transpose}{^\mathrm{T}}

\def \path{\bp C}

\newcommand{\bfm}{{\mathbf{m}}}

\newcommand{\bfx}{{\mathbf{x}}}
\newcommand{\bfy}{{\mathbf{y}}}
\newcommand{\bfz}{{\mathbf{z}}}

\newcommand{\bfK}{{\mathbf{K}}}

\newcommand{\bfM}{{\mathbf{M}}}

\newcommand{\bfQ}{{\mathbf{Q}}}

\newcommand{\bfV}{{\mathbf{V}}}

\newcommand{\bfZ}{{\mathbf{Z}}}

\newcommand{\calL}{{\mathcal{L}}}
\newcommand{\calM}{{\mathcal{M}}}

\newcommand{\bbR}{{\mathbb{R}}}

\title{Variable Attention Masking for Configurable Transformer Transducer Speech Recognition}

\name{\begin{tabular}{c} Pawel Swietojanski, Stefan Braun, Dogan Can, Thiago Fraga da Silva, Arnab Ghoshal \\
      Takaaki Hori, Roger Hsiao, Henry Mason, Erik McDermott \\ 
      Honza Silovsky, Ruchir Travadi, Xiaodan Zhuang \end{tabular}
}

\address{Apple}

\begin{document}
\ninept
\maketitle

\begin{abstract}
This work studies the use of attention masking in transformer transducer based speech recognition for building a single configurable model for different deployment scenarios. We present a comprehensive set of experiments comparing fixed masking, where the same attention mask is applied at every frame, with chunked masking, where the attention mask for each frame is determined by chunk boundaries, in terms of recognition accuracy and latency. We then explore the use of variable masking, where the attention masks are sampled from a target distribution at training time, to build models that can work in different configurations. Finally, we investigate how a single configurable model can be used to perform both first pass streaming recognition and second pass acoustic rescoring. Experiments show that chunked masking achieves a better accuracy vs latency trade-off compared to fixed masking, both with and without FastEmit. We also show that variable masking improves the accuracy by up to 8\% relative in the acoustic re-scoring scenario.
\end{abstract}


\section{Introduction}

Many applications of automatic speech recognition (ASR) require both low latency and high accuracy, goals that are not easily attained within a single system. Historically, these were achieved by building separate models oriented at either online or offline ASR. In some deployments, the predictions of a model characterized by good streaming properties, but sub-optimal accuracy, could be refined by a different model with worse latency profile, but higher accuracy.

End-to-end models~\cite{Li2022RecentAI}, in particular those utilizing transformers\cite{Vaswani2017,Zhang2020} as a back-bone, can be trained to operate in different configurations that gracefully trade accuracy for latency or compute gains. Transformer-based models are particularly suitable for such applications thanks to their combined use of self-attention and masking -- components that jointly provide direct control over the model's access to past, present and future information when making localized predictions. When combined with frame-synchronous modeling approaches, such as the neural transducer~\cite{graves2012sequence} or connectionist temporal classification (CTC) \cite{Graves2006icml}, transformer attention masking can be used to exactly mimic the target inference conditions during training and thereby acquire better control over the accuracy vs latency trade-off induced by the model topology.

An example of a neural transducer model with low partial latency, but still benefiting from access to larger context, is the stacked encoder approach \cite{Narayanan2021, Mahadeokar2022}, in which data is encoded using cascaded encoders - an encoder operating in causal mode, followed by a stacked non-causal encoder operating on the causal encoder outputs and optionally updating the causal encoder predictions. Though jointly learned, each encoder has its own set of parameters. 

Here, we build on an alternative approach, in which transformer-transducer (TT) models are trained with variable masking~\cite{Anshuman2020}. In this approach, different masking patterns are applied to the acoustic encoder during training. This exposes the encoder to varying amounts of future context, allowing the trained model to work in both streaming and non-streaming modes, while sharing parameters between the two modes. There are a number of works in recent literature investigating similar ideas. \cite{yu2020dual, Moritz2021} focused on learning causal and non-causal convolution filters in the context of ``dual-mode'' transducers with a conformer-based acoustic encoder. Doutre et al. \cite{Doutre2020} explored building multi-mode models with distillation of full-context models to streaming variants. Audhkhasi et al. \cite{Audhkhasi2021} proposed a mixture of softmaxes approach to computing the attention probability density function, where the support for mixture components may be chosen to cover different modeling contexts. In contrast, following~\cite{Anshuman2020}, we adopt a single softmax attention scheme, with masking acting as a form of conditional execution governed by sampled choices.

We also build on recent work describing attention-based causal chunking \cite{Shi2021EmformerEM,Chen2021}. In this approach, a chunk-aware attention mask allows access to past and future frames within a pre-defined chunk, but otherwise disallows information flow between chunks. The trained model can then be treated as causal at the chunk level. Here, we extend the causal chunking approach approach to variable causal chunking, thus allowing the model to work in both streaming and global modes. A similar extension was recently also made for CTC models\cite{Zhuoyuan2021}. Further, we relax the constraints on information flow by allowing the model to access distant past information at each layer in streaming mode. 

There are other studies aimed at localizing self-attention for streaming applications, not necessarily via masking \cite{huang2020conv,moritz2020streaming,tian2020synchronous, tsunoo2019transformer,zhang2020streaming}. These works do not address the issue of inference-time configurability, and thus do not directly lend themselves to unified two pass applications.

Contributions of this work include:
\begin{itemize}
\item Systematic comparison of the accuracy and latency impact of transformer masking strategies ~\cite{Anshuman2020,Shi2021EmformerEM,Chen2021} and their variable masking variants within the TT framework. 
\item Evaluation of variable masking as a strategy for inference-time configurability.
\item Evaluation of variable masking for two-pass acoustic re-scoring in the large-scale data regime.
\end{itemize}

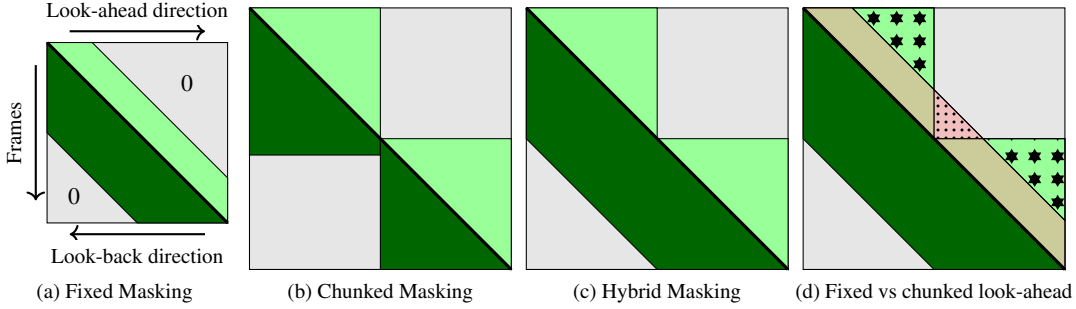
\begin{figure*}[ht]
    \centering
\subfloat[Fixed Masking] { \label{fig:mfixed}
\begin{tikzpicture}[scale=0.6]
    \draw[thick,->] (0.5,4.25) -- node[above=2pt,fill=white] {\footnotesize Look-ahead direction} (3.5,4.25);
    \draw[thick,->] (3.5,-0.25) -- node[below=2pt,fill=white] {\footnotesize Look-back direction} (0.5,-0.25);
    \draw[thick,->] (-0.25,3.5) -- node[left=2pt,fill=white] {\rotatebox{90}{\footnotesize Frames}} (-0.25,0.5);
    \filldraw[gray!20, draw=black] (0,0) rectangle (4,4);
    \filldraw[green!40!white, draw=black] (0,4) -- (1,4) -- (4,1) -- (4,0) -- (0,4);
    \filldraw[green!40!black, draw=black] (0,4) -- (0,2) -- (2,0) -- (4,0) -- (0,4);
    \draw[black,very thick] (0,4) -- (4,0);
    \node[draw=none,align=left] at (3.1, 3.1) {0};
    \node[draw=none,align=left] at (0.6,0.6) {0};
\end{tikzpicture}
}
\subfloat[Chunked Masking] { \label{fig:mchunked}
\begin{tikzpicture}[scale=0.87]
    \filldraw[gray!20, draw=black] (0,0) rectangle (4,4);
    \filldraw[green!40!white, draw=black] (0,4) -- (2,4) -- (2,2) -- (0,4);
    \filldraw[green!40!white, draw=black] (2,2) -- (4,2) -- (4,0) -- (2,2);
    \filldraw[green!40!black, draw=black] (0,4) -- (0,1.75) -- (2,1.75) -- (2,2) -- (0,4);
    \filldraw[green!40!black, draw=black] (2,2) -- (4,0) -- (2,0) -- (2,2);
    \draw[black,very thick] (0,4) -- (4,0);
\end{tikzpicture}
}
\subfloat[Hybrid Masking] { \label{fig:mhybrid}
\begin{tikzpicture}[scale=0.87]
    \filldraw[gray!20, draw=black] (0,0) rectangle (4,4);
    \filldraw[green!40!white, draw=black] (0,4) -- (2,4) -- (2,2) -- (0,4);
    \filldraw[green!40!white, draw=black] (2,2) -- (4,2) -- (4,0) -- (2,2);
    \filldraw[green!40!black, draw=black] (0,4) -- (0,2) -- (2,0) -- (4,0) -- (0,4);
    \draw[black,very thick] (0,4) -- (4,0);
\end{tikzpicture}
}
\subfloat[Fixed vs chunked look-ahead] { \label{fig:compromise}
\begin{tikzpicture}[scale=0.87]
    \filldraw[gray!20, draw=black] (0,0) rectangle (4,4);
    \filldraw[green!40!white, draw=black] (0,4) -- (2,4) -- (2,2) -- (0,4);
    \filldraw[green!40!white, draw=black] (2,2) -- (4,2) -- (4,0) -- (2,2);
    \filldraw[fill=red!40!white, fill opacity=0.5] (0,4) -- (0.75,4) -- (4,0.75) -- (4,0) -- (0,4);
    \filldraw[green!40!black, draw=black] (0,4) -- (0,2) -- (2,0) -- (4,0) -- (0,4);
    \draw[pattern={sixpointed stars}] (0.75,4) -- (2,4) -- (2,2.75) -- (0.75,4);
    \draw[pattern={dots}] (2,2.75) -- (2.75,2) -- (2,2) -- (2, 2.75);
    \draw[pattern={sixpointed stars}] (2.75, 2) -- (4,2) -- (4,0.75) -- (2.75,2);
    \draw[black,very thick] (0,4) -- (4,0);
\end{tikzpicture}
}
    \caption{Illustration of attention masking in transformers. \figref{fig:mfixed} depicts a standard fixed look-ahead/look-back mode \cite{Zhang2020}. Each row shows the current frame (on the diagonal) and its past (lower triangular) and future (upper triangular) context. \figref{fig:mchunked} shows causal chunking as in \cite{Shi2021EmformerEM, Chen2021}. \figref{fig:mhybrid} illustrates a hybrid approach, where the future direction benefits from both a non-zero look-ahead and causality across chunk boundaries.
    \figref{fig:compromise} illustrates the trade-offs from look-ahead type between fixed and chunked masking. Star-annotated areas depict the additional look-ahead that chunked masking can use without impacting look-ahead induced latency. The dot-annotated area depicts additional look-ahead enabled by fixed look-ahead, at the expense of some latency. Given the same look-ahead (latency) budget, all transformer layers can benefit from non-causality within chunks, contrary to fixed masking where only few layers are allowed to be non-causal.
    }
    \label{fig:masking}
\end{figure*}

\section {Transducers, Transformers, Attention and Masking}
\subsection{Transformer Transducer}
In this work we use ASR systems based on the TT architecture\cite{Zhang2020,Anshuman2020}, trained with the RNN-T loss \cite{graves2012sequence}. 
The predictive distribution of the transducer model is defined as:
\begin{equation}
        P( \bfy_{u+1}|\bfx, t, \;\bfy_{1:u}) = \mbox{Softmax} \left ( \mbox{Linear }( \mbox{tanh} ( \mbox{Joint} ) ) \right ),
\end{equation}
\begin{equation} \label{eq:joiner}
 \begin{split}
    \mbox{Joint} = & \; \mbox{Linear}(\mbox{AcousticEncoder}(\bfx, t)) \\
                   & + \mbox{Linear}(\mbox{LabelEncoder}(\bfy_{1:u})),
 \end{split}
\end{equation}
where $\bfx$ is the audio sequence; $\mbox{AcousticEncoder}(\bfx, t)$ is the acoustic encoder output at time $t$; $\bfy$ is the label sequence; $\mbox{LabelEncoder}(\bfy_{1:u})$ is the label encoder output given the previous non-blank tokens $\bfy_{1:u}$.
This work uses standard transformer layers \cite{Vaswani2017} for label encoder and convolution-augmented transformer (Conformer) \cite{Gulati2020} layers for acoustic encoder. The Linear function is a single affine transform.

\vspace{-0.1cm}
\subsection{Transformer Attention Masking}

Different from RNNs, transformers allow precise control of the neighborhood information at each step. For example, 
 at each time $t$, $\mbox{AcousticEncoder}(\bfx, t)$ in Eq. \eqref{eq:joiner} may be derived from an arbitrary subset of features in $\bfx$, as defined by the masking strategy implemented in the self-attention layers \cite{Vaswani2017}.
Given the attention input $\bfZ = (\bfz_1, \ldots, \bfz_{L_z})$, $\bfz_t \in \bbR^{d_z}$ self-attention computes: 
\begin{equation}
\bfQ=f^q(\bfZ), \; \bfK=f^k(\bfZ), \; \bfV=f^v(\bfZ),
\end{equation}
\begin{equation}
    \mbox{Att}(\bfQ, \bfK, \bfV) = \mbox{softmax} \left (\alpha \calM(\bfQ\bfK\transpose) \right ) \bfV\transpose,
\end{equation}
where $\alpha = 1/\sqrt{d}$ is the scaling factor dependent on the attention dimension. Vaswani et al. \cite{Vaswani2017} implemented $f^*(\bfZ)$ mappings as linear transforms. 
However, this work uses asymmetric relative positional encoding~\cite{pham2020relative} in acoustic encoder and relative positional encoding~\cite{dai2019transformerxl} for label encoder. The mask $\calM\in{\{0,1\}^{L_z\cdot L_k}}$
determines the access of each transformer frame to its past and future contexts \cite{Li2022RecentAI}. We consider several masking patterns as depicted in \figref{fig:masking}, in particular we will refer to these as i) fixed masking \cite{Zhang2020} and ii) chunked masking \cite{Shi2021EmformerEM, Chen2021}. 

The masking pattern defines the streaming properties of the model.
In low latency recognition scenarios, average acoustic encoder look-ahead should be smaller than the target partial latency and encoder outputs should be produced as soon as possible. However, computing outputs one by one significantly increases the cost of computation, e.g. power use, on parallel hardware. Hence, it is important to balance the latency with compute by chunking input frames and encoding each chunk as a batch. Generally speaking, the larger the chunk size the lower the compute cost. Exploiting this fact, \cite{Shi2021EmformerEM, Chen2021} proposed a masking strategy that uses maximal valid context within a chunk, and then restricts the model to be causal across chunk boundaries, as illustrated in \figref{fig:mchunked}. Chunked masking offers better average look-ahead when compared to fixed masking and allows to benefit from look-ahead at each layer. The compromise between fixed and chunked masking is illustrated in \figref{fig:compromise}. 
In this work, we also investigate a hybrid mode (\figref{fig:mhybrid}), where we use causal chunking for future context, and fixed masking for past context, 
which allows each layer to access distant past frames.

\subsection{Variable masking and its applications} \label{ssec:vm}

We extend each of the variants presented in \figref{fig:masking} to variable masking scenario, where during training acoustic and/or label encoders are exposed to multiple masking configurations so the model can learn to integrate different amounts of past and future context, and thus be deployed in desired operating conditions during inference. From training perspective, this can be seen as an additional conditioning variable~\ie:
\begin{equation}
\calL_{\texttt{TT}} = -\log P( \bfy|\bfx; \boldsymbol\theta_{\texttt{TT}}, \bfm\sim \bfM),
\end{equation}
\noindent where $\bfm = (\bfm_{\texttt{AE}}, \bfm_{\texttt{LE}})$ is a sample from $\bfM$, a predefined set of allowed mask configurations. $\bfm_{\texttt{AE}}$ is the acoustic encoder mask specifying the look-back and look-ahead frames (or chunk sizes) at each transformer layer. The label encoder mask $\bfm_{\texttt{LE}}$ is strictly causal and only specifies look-back frames. Similar conditioning on external variables was previously used for speaker adaptive training\cite{bell2020adaptation}.

One advantage of variable masking models is their ability to obtain encodings for different amounts of context. One can use this in Eq. \eqref{eq:joiner} to re-compute the $\mbox{Joint}$ outputs in second pass with a wide chunk (\texttt{WC}) acoustic encoder, while re-using cached label encodings:
 \begin{equation} \label{eq:joiner_rescoring}
 \begin{split}
    \mbox{Joint} = & \; \mbox{Linear} ( \mbox{AcousticEncoder}_{\texttt{WC}}(\bfx, t)) \\
                   & + \mbox{Linear}(\mbox{LabelEncoder}(\bfy_{1:U})).
 \end{split}
\end{equation}
Note that the $\mbox{AcousticEncoder}_{\texttt{WC}}(\bfx, t)$ encodings are obtained with the same acoustic encoder weights, and different mask configuration. $\mbox{LabelEncoder}(\bfy_{1:U})$ encodings are cached and reused from  Eq.~\eqref{eq:joiner}. 

\section {Experiments}

We carry out experiments on large-scale in-house data comprising of utterances from dictation and assistant tasks. We use both semi-supervised and supervised data. Semi-supervised transcripts were automatically obtained from an auxiliary transcription model applied to around 600,000 hours of randomized and anonymized acoustic data. The supervised data consists of 7,000 hours of US English data. All systems in this work were trained for 3.4M updates, each update using gradients accumulated over 8192 utterances. We use SyncSGD with exponentially decaying learning schedule. The models are evaluated on a test set that is around 60 hours long, with utterance duration varying from a few seconds to around 60 seconds. 

Our base TT models operate on 6 times decimated acoustic features by means of convolution and frame concatenation based down-sampling. The acoustic encoder makes use of 12 causal Conformer blocks. For streaming applications with variable masking, we found it crucial for training stability to use layer normalization\cite{ba2016_layernorm} rather than batch-level normalization\cite{ioffe15_batchnorm} as in the original Conformer formulation\cite{Gulati2020}. The label encoder is implemented as a standard 6 layer transformer. In total, all TT models shown have 101M parameters. 
In the following, we refer to this model as a CCT transducer, short for \textbf{C}onvolutional downsampling, \textbf{C}onformer acoustic encoder and \textbf{T}ransformer label encoder. Depending on masking configuration, we additionally annotate CCT with one of the following prefixes, \{f,vf,c,vc\}, denoting fixed context-expansion (f) look-aheads / look-backs, variable context-expansion masking (vf), chunking (c) and variable-chunking (vc) CCT models, respectively.

\begin{figure}
    \centering
    \includegraphics[scale=0.55]{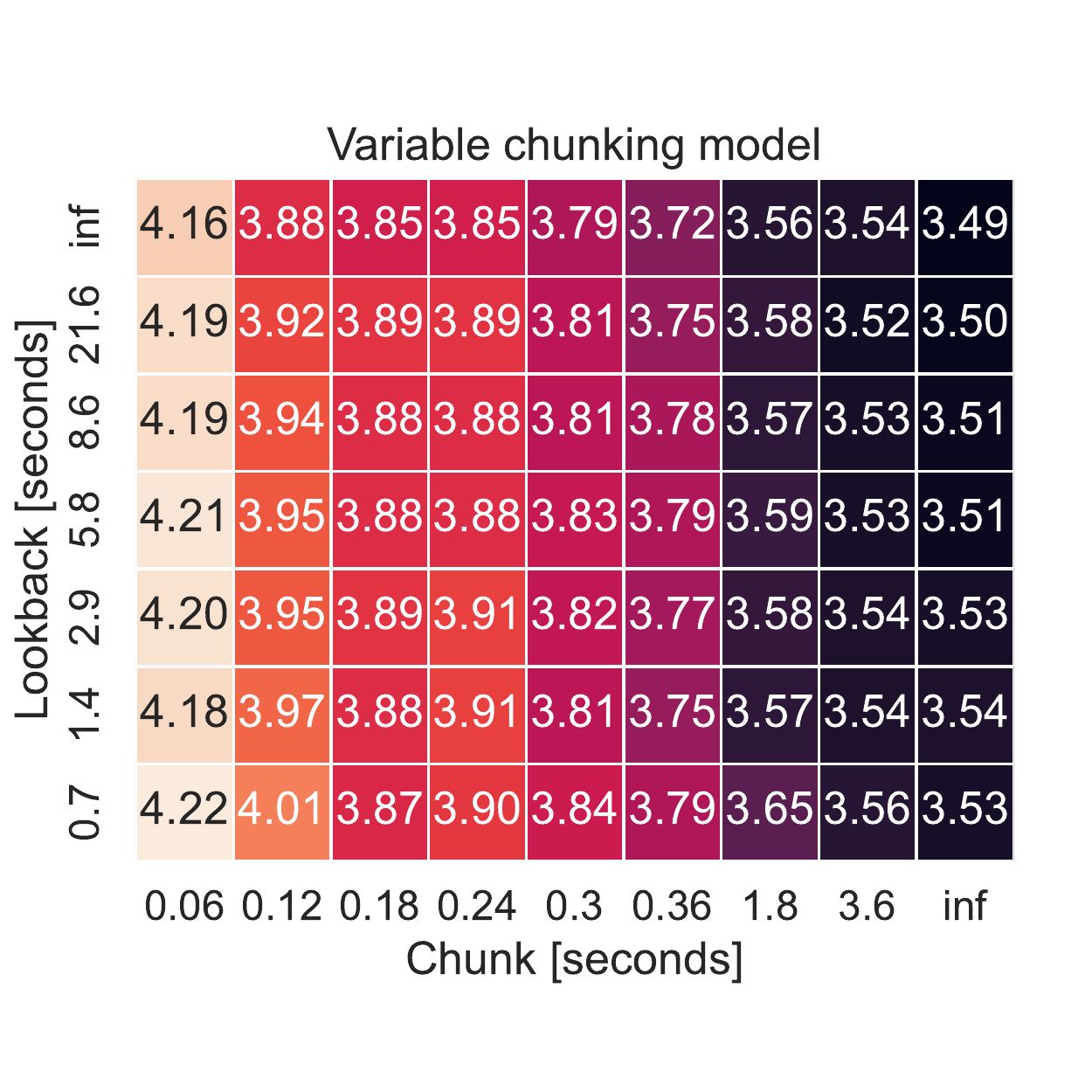}
    \vspace{-1cm}
    \caption{WERs for variable chunk masking models and a number of different masks settings used in audio encoder for the left and right context. Same trends were observed with fixed variable masking.
    }
    \label{fig:chunked_decodes}
    \vspace{-0.4cm}
\end{figure}

\subsection {Single-masking baselines}

First block in Table~\ref{tab:masking_modes} reports WERs for different masking modes outlined in~\figref{fig:masking}. The models were configured to a similar streaming setting consuming 240ms of audio data (or 4 transformer frames after 6x decimation). Fixed masking on both past and future (\figref{fig:mfixed}) is comparable accuracy-wise to a hybrid approach that uses causal chunking combined with context-expansion (fixed) look-backs (\figref{fig:mhybrid}). Causal chunking as depicted in \figref{fig:mchunked} performed about 0.1\% worse than hybrid chunking; the difference disappears if one can trade-off larger chunks for latency. As we will show later in Section \ref{ssec:partial_latencies}, causal chunking offers better streaming properties wrt partial latencies; it also simplifies decoding and saves computation by not requiring the availability of frames from future chunks.

\begin{table}[!htbp]
    \centering
    \small
    \begin{tabular}{c|c|c|c|c|c}
         \multicolumn{2}{p{2cm}|}{\centering Left Context} & \multicolumn{2}{p{2cm}|}{\centering Right Context} &  \multicolumn{2}{p{2cm}}{\centering WER [\%]} \\ \hline
         Type   & \#Masks        & Type  & \#Masks  & Stream. & Global           \\ \hline\hline
         Chunk  & 1         & Chunk & 1             &  3.75  & -       \\ 
         Fixed  & 1         & Fixed & 1             &  3.61  & -    \\
         Fixed  & 1         & Chunk & 1             &  3.66  & -     \\ \hline
         Fixed  & 1         & Chunk & 2             &  3.66  & 3.4 \\
         Fixed  & 1         & Chunk & 5             &  3.71  & 3.39  \\
         Fixed  & 1         & Chunk & 9             &  3.69  & 3.46  \\
         Fixed  & 2         & Chunk & 9             &  3.7   & 3.44 \\
         Fixed  & 7         & Chunk & 9             &  3.88  & 3.5  \\
    \end{tabular}
    \caption{WERs for different masking variants in \figref{fig:masking}.}
    \label{tab:masking_modes}
    \vspace{-0.5cm}
\end{table}

\subsection {Variable masking}

In this section we expand single-masking systems to training with variable masking configurations. We report results for Fixed-Chunk systems from first portion of Table~\ref{tab:masking_modes}, as Fixed-Fixed follows a very similar pattern. We trained models with a variety of masking settings for chunks and look-backs (LB) in acoustic encoder. We also experimented with varying the left context of label encoders; however, this was found to have a negligible impact on accuracy, in line with other works on limited context label encoders with transducers \cite{ghodsi2020, sainath21_interspeech}. We sampled the mask settings independently from each other during training, allowing the model to be arbitrarily configured to any combination of them at inference. During training, the masks are sampled at batch level\footnote{Per-utterance mask sampling offered similar WERs.}, and applied consistently across all transformer layers. 
A decode of the model from the last row in Table~\ref{tab:masking_modes} for all its operating settings is shown in \figref{fig:chunked_decodes}. The model was trained with 9 different chunk configurations, ranging from fully causal (60ms audio chunk, or 1 transformer frame) to full future context. 
Likewise, we trained with 7 different LB configurations, spanning the history from last 720ms to the full past context. It is interesting to observe how little past context is needed to obtain reasonable results, i.e. the most constrained model configuration with only the current frame and 720ms of past audio obtains 4.22\% WER, and how little WER improvement full past context brings (less than 0.1\% abs.). A similar trend can be observed for different chunk sizes, though increasing the chunk size (or the look-ahead) has a much larger impact on accuracy. Note that the test set includes audio up to 60s, and when testing the models on different long-form audio test sets (results not reported), we did not observe any dramatic increases in WER. This suggests that limiting the context via masking could be a simple but effective alternative to more sophisticated approaches preventing self-attention from collapse on long speech sequences \cite{Likhomanenko2021_CAPE}.
Since a large number of LB settings seems to hurt performance (compare the last two rows in Table~\ref{tab:masking_modes}), in the remainder we will do the analyses on the model trained with 2 LB and 9 chunk size settings. As the default inference setting, we use 8.6s of LB and a 240ms long chunk.

\subsection {Partial Result Word Latency} \label{ssec:partial_latencies}

Recognizer responsiveness is paramount in interactive recognition applications like dictation. Partial Result Word Latency (PRWL) is defined as the delay between when a word is fully spoken and when it is presented to the user. To compute timestamps of each word in our test set, we use a frame synchronous hybrid phonetic HMM-CNN recognizer to produce a 10ms frame time-aligned transcript. Every partial result of the transducer decoder is recorded along with the amount of audio consumed by the acoustic encoder to compute the \textit{first seen timestamp} of each word.

There are several techniques to encourage end-to-end ASR models to prefer lower latency predictions. One approach is to constrain the frame alignment during training to ground truth time alignment \cite{Senior2015}. These constraints do limit the model's ability to learn certain alignments, which impacts accuracy. Another approach, specific to the transducer loss function, is FastEmit (FE) \cite{yu2021_fastemit}. FastEmit has the advantage of not requiring any alignment training data, although it too can affect accuracy. A novel approach 
is self-alignment \cite{Kim2021_selfalign}, which uses Viterbi forced-alignment instead of external alignment.

\begin{figure}
    \centering
    \includegraphics[scale=0.5]{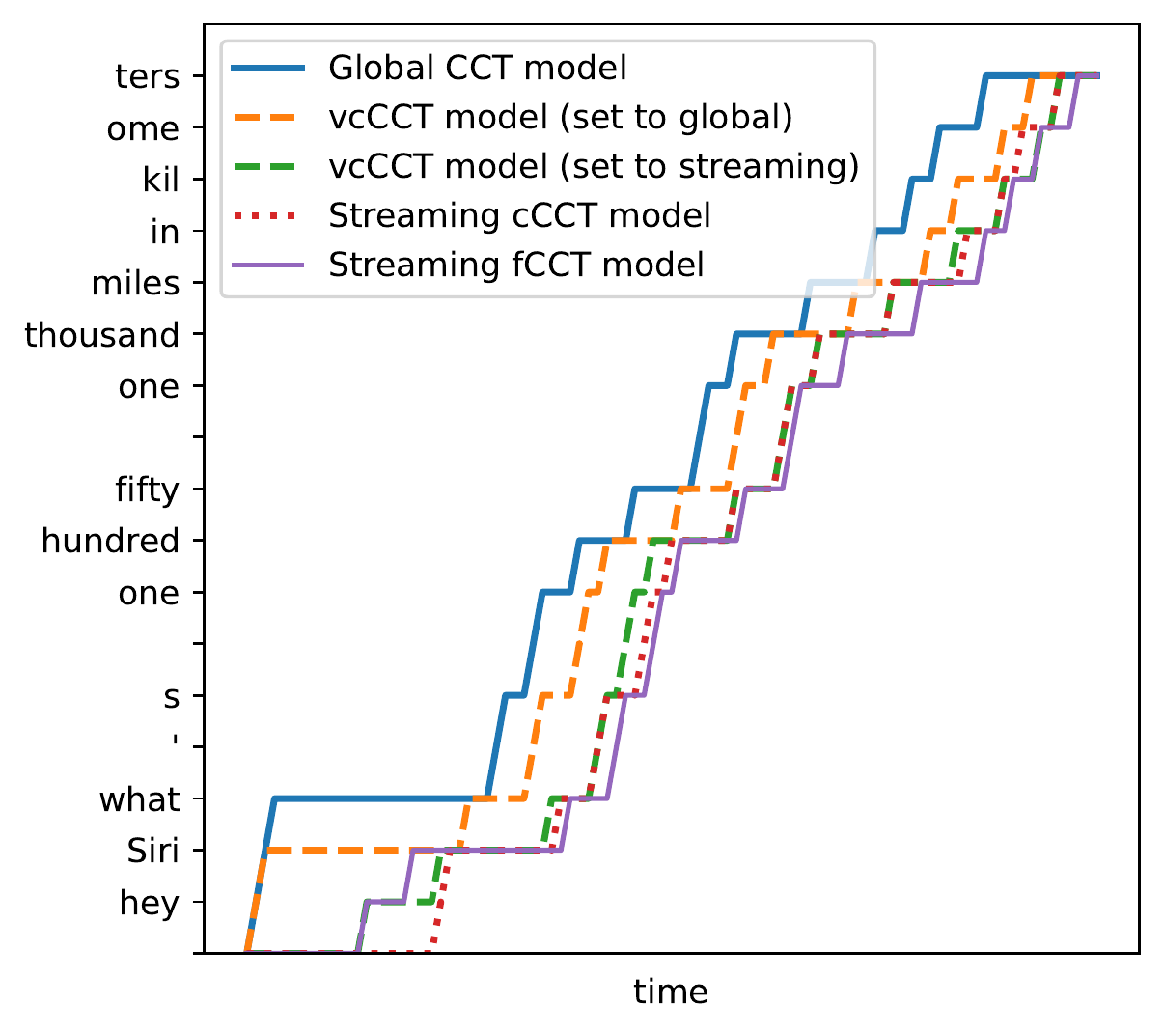}
    \vspace{-0.5cm}
    \caption{Token emission alignments for different models.}
    \label{fig:alignments}
    \vspace{-0.5cm}
\end{figure}

PRWL results for selected models are reported in Table~\ref{tab:latencies}.  Chunked masking reduces PRWL by around 45\% relative when compared to fixed masking models. For both fixed masking (\{f,vf\}-CCT) and chunking configurations (\{c,vc\}-CCT), we find FastEmit to be highly effective, as expected. Chunked models can afford less aggressive FE setting, usually resulting in better accuracy for the same latency constraints (\ie compare vfCCT and vcCCT models, where the latter obtains 45\% lower PWRL for the same FE constant).
~\figref{fig:alignments} shows example token emission delays of an FE-enabled vcCCT configured to streaming and global modes, as well as single-mask variants from Table~\ref{tab:latencies}.
Variable chunking combined with latency penalizing loss improves PRWL, and decreases token emission delay. As a function of chunk size, token emission latency decreases for larger chunks, around 40ms on average between models configured to 60ms vs 240ms chunks. 

We also investigated if the amount of look-back affects token emission delay by measuring the system from ~\figref{fig:chunked_decodes} configured to 8.6s and 720ms LB settings, and we did not find a significant effect.

Chunking, due to batched computation, brings large gains in real-time factor (RTF) or compute-induced latency. We do not report RTFs in this work as they were in-depth studied in \cite{Shi2021EmformerEM, Chen2021}.

\begin{table}[]
    \centering
    \small
    \begin{tabular}{l|c|c}
         System &  PRWL [ms] & WER [\%] \\ \hline\hline
         fCCT      & 835 & 3.56   \\
         ~+FastEmit & 700 & 3.61   \\
         vfCCT     & 855 & 3.68   \\
         ~+FastEmit & 710 & 3.71   \\
         cCCT      & 453   & 3.62  \\
         ~+FastEmit & 409 & 3.66  \\ 
         vcCCT     & 473 & 3.56 \\
         ~+FastEmit & 389 & 3.66  \\ 
    \end{tabular}
    \caption{Latency metrics for selected systems. 
    }
    \label{tab:latencies}
    \vspace{-0.2cm}
\end{table}

\subsection{Acoustic Rescoring}

\begin{figure}
    \centering
    \includegraphics[scale=0.55]{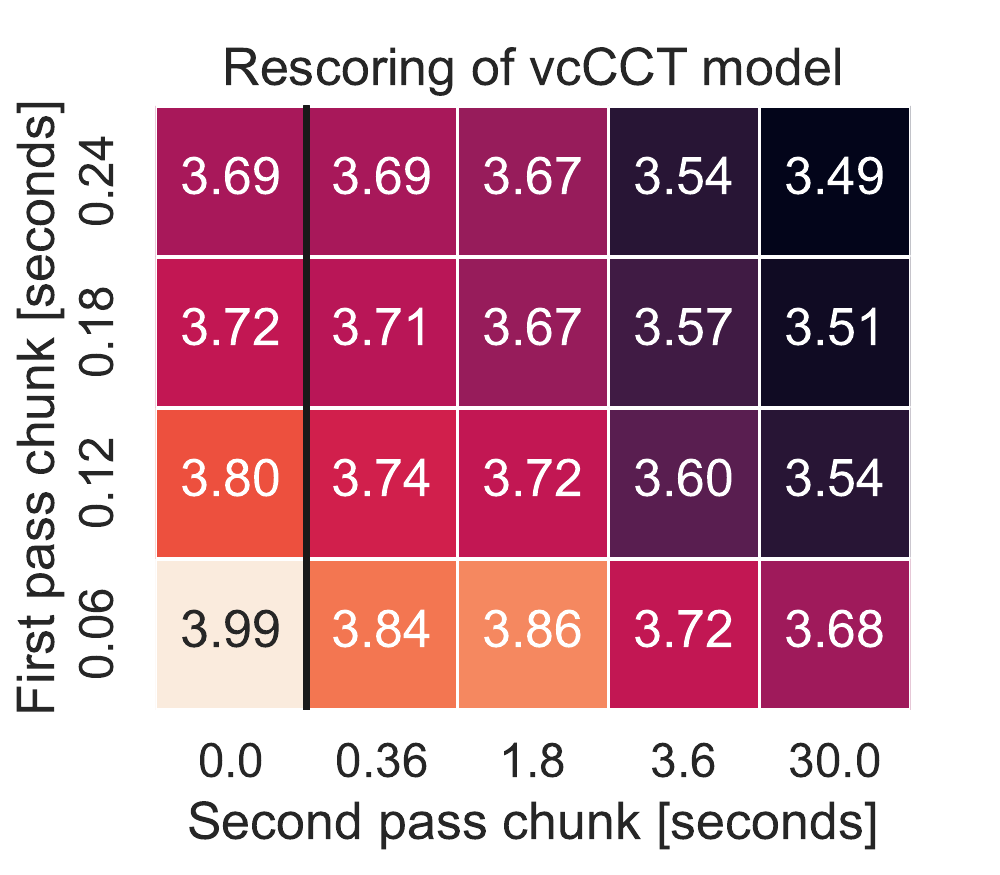}
    \vspace{-0.25cm}
    \caption{WERs obtained with acoustic re-scoring of narrow-chunk first-pass models with several wider-chunk settings. The left-most column depicts WERs for first-pass only systems.}
    \label{fig:rescoring}
\end{figure}

Variable masking models allow for an easy application to acoustic re-scoring, as they do not require auxiliary models or distinct parameters to capture different amounts of context. In particular, in this work we re-score by recomputing acoustic embeddings with an acoustic encoder configured to capture larger amounts of audio context as described in Section~\ref{ssec:vm}, and cached label encodings (thus avoiding an expensive auto-regressive recomputation). Results are depicted in \figref{fig:rescoring}. We obtain first-pass 10-best lists from a vcCCT operating on {60, 120, 180 or 240}ms long chunks, and then re-score with AE configured to 0.36, 1.8, 3.6, and 30 seconds chunks. Note that the 60ms setting is effectively a causal model, where the chunk is just one transformer frame. 
Depending on chunking configuration, we obtain up to 8\% relative WER reduction from 2nd pass re-scoring. For comparison, decodes with larger chunks gave WERs of 3.52, 3.49 and 3.46 for chunk sizes of 1.8, 3.6 and 30 seconds, respectively; re-scoring is thus on average within 0.1\% abs. WER.

\section{Conclusions}

We have shown that variable masking transformer transducer models are a viable choice for two-pass acoustic speech recognition pipelines, without the necessity of maintaining additional parameters in the form of stacked encoders or additional re-scoring modules such as attention encoder-decoder. We have investigated how different masking strategies impact partial latency, finding chunking to be twice as good when compared to fixed masking and similar accuracy constraints.
Finally, we carried out an extensive experimental evaluation of variable masking strategies in a large-scale data regime, including their application to two-pass speech recognition. 

{
\small 
\section {Acknowledgements}
We thank Vijay Peddinti, Tatiana Likhomanenko, Sicheng Wang and Russ Webb for useful suggestions on this work.
}

\newpage
\ninept
\bibliographystyle{IEEEBib}
\bibliography{sample}

\end{document}